\documentstyle[prl,aps,epsfig,twocolumn]{revtex}
\newcommand{\be}{\begin{equation}}
\newcommand{\ee}{\end{equation}}
\newcommand{\bea}{\begin{eqnarray}}
\newcommand{\eea}{\end{eqnarray}}
\newcommand{\eq}[1]{Eq.~(\ref{eq:#1})}
\newcommand{\eqs}[2]{Eqs.~(\ref{eq:#1}) and (\ref{eq:#2})}
\newcommand{\B}[1]{|\Phi_#1\>}
\def\<{\langle}
\def\>{\rangle}

\def\lpm{ \left(\rule{0pt}{2.1ex}\right. }
\def\rpm{ \left.\rule{0pt}{2.1ex}\right) }

\def\dag{\dagger}
\def\ot{\otimes}

\def\non{\nonumber}
\def\os2{{1 \over \sqrt{2}}}
\def\s{\sigma}
\def\v1{\vspace*{0.7ex}}

\begin{document}
\title{Two-qubit Projective Measurements are Universal for Quantum 
Computation}
\author{Debbie W. Leung} 
\address{IBM T.J.~Watson Research Center, P.O.~Box 218, 
Yorktown Heights, NY 10598, USA \\
Institute of Theoretical Physics, University of California, Santa Barbara, 
CA 93106-4030, USA\\[1.5ex]}
\date{April 04, 2002}
\maketitle
%
\begin{abstract}
Nielsen [quant-ph/0108020] showed that universal quantum computation
is possible given quantum memory and the ability to perform projective
measurements on up to $4$-qubits.
We describe an improved method that requires only $2$-qubit
measurements, which are both sufficient and necessary.
We present a method to partially collapse the $C_k$-hierarchy in the
indirect construction of unitary gates [Gottesman and Chuang, Nature,
{\bf 402} 309 (1999)], and apply the method to find discrete universal
sets of $2$-qubit measurements.
\end{abstract}

\vspace*{2ex}


The standard model of quantum computation~\cite{DiVincenzo95a}
requires a well defined and isolated Hilbert space.
Universal computation further requires the ability (1) to prepare
a fiducial initial state,
(2) to implement a {\em universal} set of gates in a quantum circuit, 
and (3) to perform strong measurements.  
The $|0\>$ state and the measurement along the computation basis
$\{|0\>,|1\>\}$ are often assumed in (1) and (3).
A set of quantum gates is universal if any unitary evolution can
be approximated to arbitrary accuracy by a circuit involving those
gates only~\cite{Universal}.
The set of all $1$-qubit gates and the controlled-{\sc not} ({\sc
cnot}) is universal, and so are some simple discrete sets of
gates.

\v1 

Other models have been built upon the standard model, so as to achieve
fault tolerance or to adapt to promising physical systems.  In these
models, only some unitary gates can be easily performed and they do
not form a universal set.
In the context of fault tolerant quantum computation, Shor pioneered a
recipe that {\em indirectly} effects the Toffoli gate using an ancilla,
measurements, and some other gates~\cite{Shor96}.
The method was generalized~\cite{Gottesman99,Zhou00} by
understanding the connection to teleportation~\cite{Bennett93}.
The generalization was applied to linear optics quantum
computation~\cite{Knill01}.
Indirect gate implementation was also studied in the context of
programmable gate arrays~\cite{Nielsen96,Vidal01g}.

\v1

More recently, Nielsen \cite{Nielsen01p} extended the above results to 
a ``measurement model'' of quantum computation that achieves the $3$
requirements of the standard model using only measurements on up to
$4$ qubits.
%
%
Fenner and Zhang \cite{Fenner01} and independently Leung and Nielsen 
\cite{Leung01mit} then showed that $3$-qubit measurements are
sufficient.
Since $2$-qubit measurements are the only means of interaction in the
measurement model, they are obviously necessary.
In this paper, we show that $2$-qubit measurements are also sufficient.
This parallels the universality of $2$-qubit gates in the standard
model~\cite{DiVincenzo95b}.
We also present simple discrete universal sets of $2$-qubit
measurement operators.  

\v1

We remark that Raussendorf and Briegel~\cite{Raussendorf01} have
proposed a very different measurement model which starts with a {\em
cluster state}~\cite{Briegel00} and uses only $1$-qubit measurements.
The cluster state is highly entangled, but can be prepared by
$4$-qubit measurements.
It remains open whether the two measurement models are related, though
there are striking similarities.

\v1

Before we present our results, we review crucial elements leading to
them.  We first review the Pauli and Clifford groups (see
\cite{Gottesman97,Nielsen00,Leung00} for example).
Let $\sigma_{1,2,3}$ or $X$, $Y$, $Z$ be the Pauli operators and
$\sigma_0$ be the $2 \times 2$ identity matrix.  The Pauli group is
generated by Pauli operators acting on each qubit.  The Clifford group
is the group of unitary operations that conjugate Pauli operators to
Pauli operators.  The Clifford group is generated by the {\sc cnot},
the phase gate $\mbox{\sc p} = e^{-i {\pi \over 4} Z}$, and the
Hadamard gate $\mbox{\sc h} = \os2 (X+Z) $.

\v1

We now review the method to perform unitary gates indirectly.  The
crucial element is teleportation~\cite{Bennett93}, in which one
transmits a qubit $|\psi\> = a |0\> + b |1\>$ using the following
circuit:
\bea
\mbox{\psfig{file=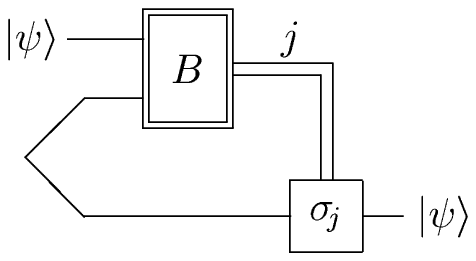,width=1.6in}}
\label{eq:teleport} 
\eea
In the circuits throughout this paper, time goes from left to right,
single lines denote qubits, double lines denote classical data,
single- and double-lined boxes respectively denote unitary gates and
measurements.  
A gate connected to a measurement box by a double line is performed
conditioned on the measurement outcome.  
Two qubits connected at the left denote a maximally entangled state 
$\B{0} = \os2 (|00\> + |11\>)$. 
The Bell measurement, labelled $B$, is along the Bell basis: 
\bea
	\B{0} & = &\os2 (|00\> + |11\>) \,,~~
	\B{3} = \os2 (|00\> - |11\>) \,,
\non
\\
	\B{1} & = & \os2 (|01\> + |10\>) \,,~~
	\B{2} = \os2 (|01\> - |10\>) \,.
\non
\eea
Equation (\ref{eq:teleport}) can be verified by rewriting the initial
state $|\psi\> \B{0}$ as
\bea
	{1 \over 2} \lpm \B{0} |\psi\> + \B{1}  (\sigma_1 |\psi\>) 
 	+ \B{2}  (\sigma_2 |\psi\>) + \B{3}  (\sigma_3 |\psi\>) \rpm .
\non
\eea
The Bell measurement on the first two qubits collapses the initial
state to one of the four terms, and conditioned on the outcome $j$, 
$\s_j$ is applied on the last qubit to recover $|\psi\>$.

\v1

Teleportation was initially proposed as a communication protocol, but
it is also useful in indirect gate constructions.
In particular, to apply a gate $U$ to a state $|\psi\>$, one can first
teleport $|\psi\>$ and then apply $U$:  
\bea
\mbox{\psfig{file=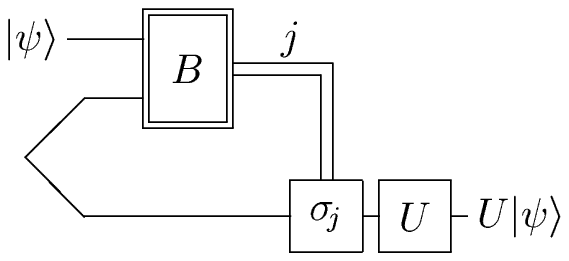,width=1.75in}}
\label{eq:homersimpson}
\eea
which implies the following circuit:
\bea
\mbox{\psfig{file=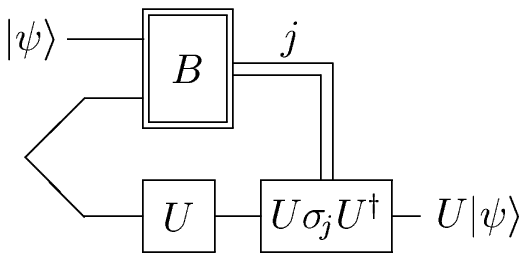,width=1.75in}}
\label{eq:heart}
\eea
Note that the gate $U \s_j U^\dagger$ is the same as the gates 
$U^\dagger$, $\s_j$, and $U$ applied in order.
Equation (\ref{eq:heart}) is a recipe for an indirect implementation
of $U$ by preparing a state $\os2 (I \ot U)(|00\>+|11\>)$ and applying
a Bell measurement and a ``correction'' $U \s_{j} U^\dag$.
It is indirect because the required ancilla $\os2 (I \ot
U)(|00\>+|11\>)$ and the correction $U \s_{j} U^\dag$ can often be
obtained without applying $U$ (see
\cite{Shor96,Boykin99,Gottesman99,Zhou00}, \cite{Nielsen96,Vidal01g},
and \cite{Knill01,Nielsen01p} in the contexts of fault-tolerance,
programmable gate arrays, and alternative computation models).
Equation (\ref{eq:heart}) can be generalized to any $n$-qubit gate.  
For instance, a $2$-qubit gate can be performed using the following 
circuit: 
\bea
\mbox{~~~~~\psfig{file=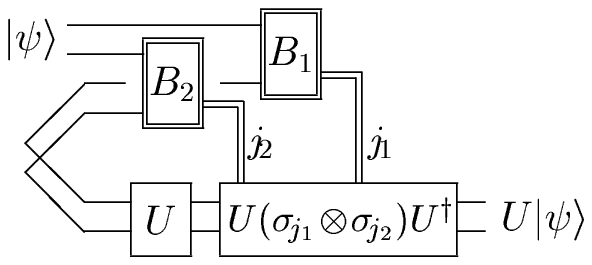,width=2.2in}}
\label{eq:heart2}
\eea

We are now ready to understand Nielsen's method to achieve the $3$
requirements in the standard model using {\em nondemolition}
projective measurements on up to $4$ qubits.
The only nontrivial requirement is to perform a universal set of gates. 
Using \eqs{heart}{heart2}, one can perform the universal set of
$1$-qubit gates and {\sc cnot} as follows.
To perform a $1$-qubit gate $U$ using \eq{heart}, it suffices to
prepare the state $(I \ot U) \B{0}$, perform the $2$-qubit Bell
measurement, and {\em sometimes} apply a correction $U \sigma_j
U^\dagger$ (when $j \neq 0$).
First, the $2$-qubit state $(I \ot U) \B{0}$ can be prepared by a
$2$-qubit projective measurement along the basis $\{(I \ot U)
\B{k}\}_k$.
Second, with probability $1/4$, $j = 0$ and no correction is needed.
Otherwise, the $1$-qubit correction gate $U \sigma_j U^\dagger$ can be
performed {\em recursively} using \eq{heart}, now with a different
ancilla, and again with probability $1/4$ no further correction is
needed.
One repeats the correction until it is no longer necessary, which
on average occurs after $4$ trials.
The {\sc cnot}, and in fact any $2$-qubit gate $U$, can be performed
analogously using \eq{heart2}.
The $4$-qubit ancilla can be prepared by a $4$-qubit measurement.
With probability ${1 \over 16}$, no correction is needed; otherwise,
another $2$-qubit correction gate is applied recursively.  On average,
$16$ trials are needed.
Nielsen also noted that ancilla preparation can be much simplified
if one replaces the construction in \eq{heart} by its variant
\bea
\mbox{~~~~~\psfig{file=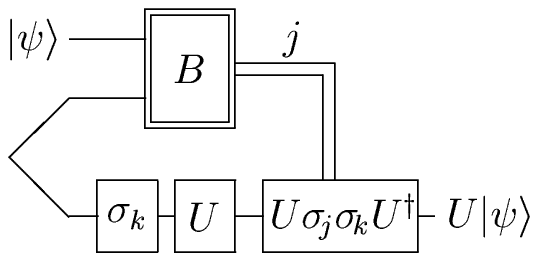,width=1.8in}}
\label{eq:heartvar}
\eea
and similarly for \eq{heart2}.  Using \eq{heartvar} instead of
\eq{heart}, the state $(I \otimes U \s_k) \B{0} = (I \otimes U) \B{k}$
makes a good ancilla for any $k$, so that a single complete
measurement is sufficient for the ancilla preparation.
Finally, we emphasize that the identity and {\sc swap} operations are
implicitly used as quantum storage and as the ability to choose which 
qubit to measure.  

\v1

To prove the universality of $2$-qubit measurements, we first make an
important observation that no $2$-qubit gate other than {\sc cnot} is
needed in the above construction: the {\sc cnot} is in the Clifford
group and all correction gates $\mbox{\sc cnot} \; (\sigma_{j_1} \ot
\sigma_{j_2}) \; \mbox{\sc cnot}$ are tensor products of
Pauli operators.
Thus only $2$-qubit measurements are used, except for preparing 
the ancilla for {\sc cnot}, which is given by
\bea
	& & \!\!\!\! \!\! |a_{\rm cn}\> 
\non
\\
	& = & {1 \over 2} (I \ot I \ot \mbox{\sc cnot})
	(|0000\> + |0101\> + |1010\> + |1111\>)
\non
\\
	& = & {1 \over 2} 
	(|0000\> + |0101\> + |1011\> + |1110\>) \,. 
\label{eq:acn}
\eea
%
%
This observation was made independently by Fenner and Zhang
\cite{Fenner01} and by Leung \cite{Leung01mit}.  In both
cases, methods to prepare $|a_{\rm cn}\>$ using $3$-qubit measurements
were found.

\v1

We now prove the universality of $2$-qubit projective measurements by
showing how they can be used to prepare $|a_{\rm cn}\>$.
For simplicity, we focus on measurement outcomes that result in
$|a_{\rm cn}\>$.  We will show later other measurement outcomes
result in equally good ancillas.
We first present the method in the state representation: \\[1.2ex]
1. Create ${1 \over 2} (|0\>+|1\>) \ot |0\> \ot (|00\>+|11\>)$
with $1$- and $2$-qubit measurements.  \\[1.2ex]
2. Apply to the 2nd and 3rd qubits the measurement with 2 projectors: 
\bea 
	& P_+ & = \B{0} \<\Phi_0| + |\Phi_1\>\<\Phi_1|
\non
\\
	& = & {1 \over 2} (|00\> \! + \! |11\>) (\<00|\!+\!\<11|) 
	+ {1 \over 2} (|01\>\!+\!|10\>) (\<01|\!+\!\<10|) 
\non
\\
	& P_- & = |\Phi_2\>\<\Phi_2| + |\Phi_3\>\<\Phi_3|
\non
\\
	& = & {1 \over 2} (|00\>\!-\!|11\>) (\<00|\!-\!\<11|) 
	+ {1 \over 2} (|01\>\!-\!|10\>) (\<01|\!-\!\<10|)
\non
\eea
When the outcome corresponds to $P_+$, the state becomes ${1 \over 2
\sqrt{2}} (|0\>+|1\>) \ot (|000\> + |011\> + |101\> + |110\>)$. \\[1.2ex]
3. Measure the parity of the 1st and 3rd qubits.  If the outcome is
even, the remaining state is given by
${1 \over 2}(|0000\> + |1011\> + |0101\> + |1110\>)$, 
which is $|a_{\rm cn}\>$. \\[1.2ex]
One can interpret steps 1 and 2 as preparing in the last $3$ qubits
the state ${1 \over 2} (|000\> + |011\> + |101\> + |110\>)$ by a {\em
partial} teleportation of $|0\>$ which picks out $\B{0} |0\> + \B{1}
X|0\>$.

\v1

We now explain the above scheme in the stabilizer
language~\cite{Gottesman97,Gottesman98}.
The stabilizer of an $n$-qubit state $|\psi\>$ is an abelian group
with $n$ generators $O_i$ such that $O_i |\psi\> = |\psi\>$.
These generators specify the state up to a phase.  
If $O |\psi\> = |\psi\>$, $U O U^\dagger (U |\psi\>) = U|\psi\>$,
therefore, when the state evolves as $|\psi\> \rightarrow
U|\psi\>$, each generator evolves as $O \rightarrow U O U^\dagger$.
Furthermore, suppose $M$ is a traceless operator with eigenvalues $\pm
1$, and it commutes or anticommutes with each generator.
If the outcomes $\pm 1$ are obtained when measuring $M$, the
generators that anticommute with $M$ evolve as $\{N_1, N_2, N_3,
\cdots\} \rightarrow \{\pm M, N_1 N_2, N_1 N_3, \cdots\}$.

\v1

The stabilizer of $\B{0}_{1,3} \ot \B{0}_{2,4}$ is generated by
$XIXI$, $ZIZI$, $IXIX$, $IZIZ$, with $\otimes$ omitted.  Let
\bea
\non  	U_{XI} & = & {\sc cnot} \, (X I) \, {\sc cnot} = X \!X \,,  
\\
\non  	U_{IX} & = & {\sc cnot} \, (I X) \, {\sc cnot} = IX \,, 
\\
\non	U_{ZI} & = & {\sc cnot} \, (Z I) \, {\sc cnot} = ZI \,, 
\\
\non  	U_{IZ} & = & {\sc cnot} \, (I Z) \, {\sc cnot} = ZZ \,. 
\eea
Then, the stabilizer of $|a_{\rm cn}\>$ is generated by: 
\bea
	\begin{array}{ll}
	XI & ~U_{XI} 
\\	ZI & ~U_{ZI} 
\\	IX & ~U_{IX} 
\\	IZ & ~U_{IZ} 
	\end{array}
~~=~~ 
	\begin{array}{ll}
	XI & ~X \! X
\\	ZI & ~ZI 
\\	IX & ~IX 
\\	IZ & ~ZZ 
	\end{array}
\label{eq:ancgen}
\eea
One can prepare a state by measuring the generators of its stabilizer.
However, any generator set for $|a_{\rm cn}\>$ contains elements of
weight $3$ (the weight is the number of nontrivial tensor components).
Our strategy is to start with initial generators of weights $1$ and
$2$ (corresponding to our initial state in step 1) and apply $2$-qubit
measurements $IX (U_{XI} U_{IX}) = IXXI$ and then $ZI U_{ZI} = ZIZI$
to {\em induce} multiplications between generators that anticommute
with the measured operator, thereby increasing the weights of the
generators.
Assuming $+1$ outcomes, the evolution is given by:
\bea
\non
	\begin{array}{ll}
	XI & ~II
\\	IZ & ~II 
\\	II & ~U_{XI} 
\\	II & ~U_{IZ} 
	\end{array}
	& \stackrel{\begin{array}{c}{\rm measure}
          \\{IX (U_{X\!I} \, U_{I\!X})}\end{array}}{\longrightarrow} & 
	\begin{array}{ll}
	XI & ~II
\\	IX & ~(U_{XI} \times U_{IX})
\\	II & ~U_{XI} 
\\	IZ & ~U_{IZ} 
	\end{array}
\\
\non
	& \stackrel{\begin{array}{c}{\rm measure}
          \\{ZI U_{Z\!I}}\end{array}}{\longrightarrow} &
	\begin{array}{ll}
	ZI & U_{ZI}
\\	XX & ~(U_{XI} \times U_{IX})
\\	XI & ~U_{XI} 
\\	IZ & ~U_{IZ} 
	\end{array}
\eea
The final set of generators is equivalent to that in \eq{ancgen}
because multiplying one generator to another does not affect the stabilizer.

\v1

So far we have focused on measurement outcomes that result in $|a_{\rm
cn}\>$.  
We now show that other outcomes result in states of the form $(\s_{k}
\ot \s_{l} \ot \mbox{\sc cnot}) \B{0}_{1,3} \ot \B{0}_{2,4} = \pm {1
\over 2} \lpm I \ot I \ot (\mbox{\sc cnot} \; \s_{k} \ot \s_{l}) \rpm
\lpm \B{0}_{1,3} \ot \B{0}_{2,4} \rpm$ which are also good ancillas 
following \eq{heartvar}.
This is because different outcomes affect the resulting states 
only by $\pm$ signs of their stabilizer generators.
Pauli operators applied to the first $2$ qubits of $|a_{\rm cn}\>$ 
also result in such signs, so that those resulting states are 
precisely $(\s_{k} \ot \s_{l} \ot II )|a_{\rm cn}\>$.
This can also be verified in the state representation.  

\v1

We now describe a variation of \eq{heart} that partially collapses the
$C_k$ hierarchy~\cite{Gottesman99} and simplifies universality study in
the measurement model.
In \eq{homersimpson}, we teleport $|\psi\>$ and then apply $U$.  We
can instead apply the gate $U$ and then teleport
$|\psi\>$~\cite{newtrick}:
\bea
\mbox{\psfig{file=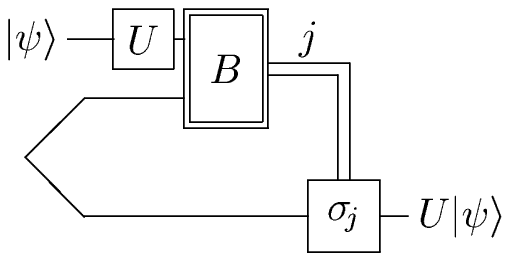,width=1.8in}}
\label{eq:gottesman}
\eea
The classical outcome in \eq{gottesman} is unchanged if $U$ and the
Bell measurement $B$ are replaced by a measurement $B_{U^\dagger_1}$
along the basis $\{ (U^\dagger \ot I) \B{j} \}_j$.
This gives a variant of \eq{heart}: 
\bea
\mbox{\psfig{file=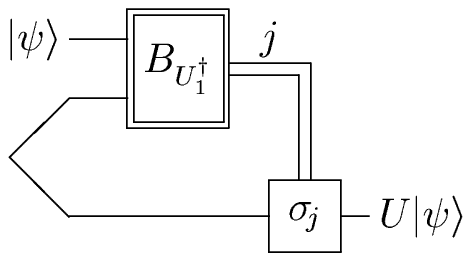,width=1.65in}}
\label{eq:bu}
\eea
Equations (\ref{eq:heart})-(\ref{eq:bu}) each uses a Bell measurement
and a special measurement, $B_{U_2}$ (along $\{(I \ot U) \B{k}\}$) for
\eq{heart} and $B_{U^\dagger_1}$ for \eq{bu}, but the functions of the
$2$ measurements, ancilla preparation and teleportation, are
interchanged.
The advantage of \eq{bu} is that the correction gate is always a Pauli
operator (instead of $U \sigma_j U^\dagger$ for \eq{heart}).
In general, if $U \in C_k$, the correction gate in \eq{bu} is in
$C_1$ while that in \eq{heart} is in $C_{k-1}$~\cite{Gottesman99,Zhou00}.

\v1

We can now apply \eqs{bu}{heart2} to find discrete universal sets of
(incomplete) $2$-qubit measurements that correspond to discrete
universal sets of gates.
It is known that the Clifford group generated by $\{${\sc cnot}, {\sc
h}, {\sc p}$\}$ together with any other gate are
universal~\cite{Gottesmanpriv}.  Thus $\{${\sc cnot}, {\sc h}, {\sc
p}, {\sc u}$\}$ is universal for any 1-qubit gate {\sc u} outside 
the Clifford group.
Applying \eqs{bu}{heart2} to this universal set, all correction gates
are Pauli operators, which require the Bell measurement or measuring
$XX$ and $ZZ$.
We also need $B_{\mbox{\sc h}^\dagger_1}$, 
$B_{\mbox{\sc u}^\dagger_1}$, and  
$B_{\mbox{\sc p}^\dagger_1}$. 
In general, $B_{U^\dagger_1}$ is a measurement of $(U^\dagger X U)
\ot X$ and $(U^\dagger Z U) \ot Z$.
Thus we need to measure $XZ$ for {\sc h},
%
%
$XY$ for {\sc p}, and $(\mbox{\sc u}^\dagger X \mbox{\sc u}) \ot
X$ and $(\mbox{\sc u}^\dagger Z \mbox{\sc u}) \ot Z$ for {\sc
u}.
Finally, we measure $X$ and $Z$ to prepare $|0\>$ and $\os2 (|0\>+|1\>)$
for $|a_{\rm cn}\>$, and for the readout.
However, $|0\>^{\otimes 3}$ can be prepared by measuring $ZZI$, $ZIZ$,
$IZZ$, and applying {\sc h} gives $\os2 (|0\>+|1\>)$.
Readout can be done by measuring $ZZ$ on the measured qubit and an
extra $|0\>$.
Thus 
\bea
	S_0 = \!\! 
	\{ XX, ZZ, XZ, XY, (\mbox{\sc u}^\dagger X \mbox{\sc u}) \ot X, 
	(\mbox{\sc u}^\dagger Z \mbox{\sc u}) \ot Z \}
\non
\eea
is universal. 
Special choices of {\sc u} can further simplify the universal set.  
For instance, 
\bea
S_1 & = \! & \{ XX, ZZ, XZ, XY, 
(\cos \theta \, Z + \sin \theta \, Y) \ot Z \}
\non
\\ 
S_2 & = \! & \{ XX, ZZ, XZ, XY, 
(\cos \theta \, X + \sin \theta \, Y) \ot X \}
\non
\\
S_3 & = \! & \{ XX, ZZ, XZ, \os2 (X+Y) \ot X \}
\non
\eea 
are universal sets of measurement operators corresponding to {\sc u}
$= e^{i {\theta \over 2} X}$, $e^{-i {\theta \over 2} Z}$, and
$e^{-i {\pi \over 8} Z}$ respectively ($\theta \neq m \pi/2$ for
$m$ an integer).
$S_3$ corresponds to the universal set of gates $\{e^{-i {\pi \over 8}
Z},$ {\sc h}, {\sc cnot}$\}$,
%
%

To conclude, we have shown how to perform universal quantum computation
using only $2$-qubit measurements, which are optimal in the number of
qubits to be jointly measured. 
We describe improved methods for performing gates by measurements, 
leading to simple discrete universal sets of $2$-qubit measurements.  
Our method requires incomplete measurement, and suggests intrinsic
differences between complete and incomplete measurements.
Finally, though experimental advantages due to \cite{Nielsen01p} and
the present work are yet to be found, alternative models for quantum
computation and their universality requirements are important for new
experimental directions and insights on what makes quantum computation
powerful.

\v1

We thank Michael Nielsen and David DiVincenzo for interesting ideas
and discussions motivating the current result, and also Charles
Bennett, Isaac Chuang, John Smolin, and Barbara Terhal for additional
helpful discussions and encouragements.
This research is supported in part by the NSA and ARDA under the US
Army Research Office, grant DAAG55-98-C-0041, and by the National
Science Foundation, grant PHY99-07949.


\begin{thebibliography}{10}
\bibitem{Leung01mit}
Leung proved that $3$-qubit measurements are universal (guest lecture, MIT
  MAS.961, Oct.~2001) because it only takes Pauli operators to correct for {\sc
  cnot}, and the ancilla for {\sc cnot} can be prepared by measuring the parity
  of the last $3$ qubits of ${1 \over 2 \sqrt{2}}
  (|00\>+|11\>)(|0\>+|1\>)(|0\>+|1\>)$. Nielsen subsequently noted that the
  ancilla for any $2$-qubit gate has stabilizers of weight $\leq 3$, and can be
  prepared by $3$-qubit measurements.
\bibitem{newtrick}
This trick has been implicitly used by Gottesman~\cite{Gottesman97} to
  construct fault tolerant Clifford group gates for general stabilizer codes.
\end{thebibliography}

\end{document}